\documentstyle[11pt]{article}

\newcommand{\xn}{x_{n}}

\newcommand{\xm}{x_{m}}

\newcommand{\kim}{ k_{1}^{\mu}}                                      
\newcommand{\kom}{ k_{0}^{\mu}}                                      
\newcommand{\ki}{ k_{1}}
\newcommand{\yn}{ Y_{n}}                                             
\newcommand{\ym}{ Y_{m}} 
\newcommand{\kn}{ k_{n}}

\newcommand{\km}{ k_{m}}

\newcommand{\kt}{ k_{2}}                                             
\newcommand{\ko}{ k_{0}}

\newcommand{\kin}{ k_{1}^{\nu}}                                      
                                      
\newcommand{\ktm}{ k_{2}^{\mu}} 
\newcommand{\ktn}{ k_{2}^{\nu}}                                      
                                      
\newcommand{\lp } {e^{ia\int _{c} k(s) \partial _{z} X(z+as) ds
 + ik_{0} X(z)}}                                                    
\newcommand{\lpp} {e^{i \int _{c} \alpha (s) 
k(s) \partial _{z} X(z+s) ds +ik_{0}X(z)}}

\newcommand{\gvk}{ e^{i\sum _{n \ge 0 }k_{n}Y_{n}(z)}}

\newcommand{\p}{\partial}                                           
\newcommand{\pp}{\partial ^{2}}                                     
\newcommand{\mup}{\partial _{\mu}}

\newcommand{\eps}{ \epsilon}                                        
\newcommand{\al}{\alpha }                                             
\newcommand{\aln}{\alpha _{n}} 
\newcommand{\tY}{\tilde Y}                                 
\newcommand{\lan}{\langle}
\newcommand{\ran}{\rangle}

\newcommand{\la}{\mbox{$ \lambda $}} 
\newcommand{\be}{\begin{equation}}
\newcommand{\br}{\begin{eqnarray}}
\newcommand{\ee}{\end{equation}} 
\newcommand{\er}{\end{eqnarray}}

\begin{document}
\renewcommand{\theequation}{\thesubsection.\arabic{equation}}

\title{
\hfill\parbox{4cm}{\normalsize IMSC/2002/01/02\\
                               hep-th/0201216}\\        
\vspace{2cm}
Loop Variables and Gauge Invariance in (Open) Bosonic String Theory.
\author{B. Sathiapalan\\ {\em Institute of Mathematical Sciences}\\
{\em Taramani}\\{\em Chennai, India 600113}\\ bala@imsc.ernet.in}}           
\maketitle     

\begin{abstract} 
We give a simplified and more complete description of the loop 
variable approach for writing down gauge invariant equations 
of motion for the fields of the open string. A simple proof 
of gauge invariance to all orders is given. In terms of loop 
variables, the interacting equations look exactly like the free 
equations, but with a loop variable depending on an extra 
parameter, thus making it a band of finite width. The arguments for
gauge invariance work exactly as in the free case. We show that these
equations are Wilsonian RG equations with a finite world-sheet cutoff 
and that in the infrared limit, equivalence with the Callan-Symanzik 
$\beta$-functions should ensure that they reproduce the on-shell
scattering amplitudes in string theory.  It is applied to the 
tachyon-photon system and the general arguments for gauge invariance 
can be easily checked to the order calculated. One can see that when 
there is a finite world sheet cutoff in place, even the U(1) 
invariance of the equations for the photon, involves massive mode 
contributions. A field redefinition involving the tachyon is required 
to get the gauge transformations of the photon into standard form.

\end{abstract}

\section{Introduction}

 The renormalization group equations (beta functions)
for the 2-dimensional action of a string in a non-trivial background
is expected to give the equations of motion for the modes of the string
[\cite{L}-\cite{T}].
This is expected to be 
true both for the closed string modes as well as open string modes.
For massless modes, which were the first to be studied,
this is relatively easy. In certain limits it can be done to all orders \cite{CANY,FT2}.
 For the tachyon also
it has been done in some detail \cite{DS,BSPT} and in some limits can be done
to all orders \cite{W2,WL,SS,KMM}. 
It is not too difficult because there are no issues of gauge invariance.
The question then arises:
How does one do this for the (interacting) massive modes?
This question has been addressed in many places, for instance in
 \cite{HLP,KPP,BGKP,W2,WL}.  
For the open string, we argue that the loop variable approach gives
an answer to this question.

At the  free level, equations were written down in \cite{BSLV}. 
A prescription for the interacting case was given in \cite{BSLV0,BSPuri}
and many details were worked out in \cite{BSLV1,BSLV2,BSWF}.
In this paper we give a simplified and more complete treatment of the  problem.
A  field redefinition at the loop variable level turns out
to simplify all the arguments in the earlier papers and gauge invariance
is much more transparent. It is easy to show that
the final sytem of equations has the
property of being gauge invariant off shell. The relation between these
equations and the equations that
produce the correct scattering amplitudes for the on-shell physical states,
is the same as that between the Wilson renormalization group equations
with finite cutoff and the Callan-Symanzik beta function. 
Thus one can expect that when one solves for the irrelevant operators
one will reproduce the on-shell scattering amplitudes. 
As an illustration we  also check the gauge invariance by explicit calculation
in the case of the tachyon-photon
system. 
This method would thus seem (at tree level) to be an alternative to BRST 
string field theory
\cite{WS,SZ,W1}. 

We have not investigated what happens at the one loop 
level. For closed strings the free equations seem to be obtainable
in this approach \cite{BSLV3}. The interactions have not been investigated.

\section{Loop Variable}

\subsection{Free Theory}

We write  the string field as a generalized Fourier transform.
\be
\Phi [X(z+s)] = \int {\cal D}k(s) \lp \Psi[k(s)]
\ee

The object $\lp$ is what is referred to as the loop variable and
can also be thought of as a collection of all the vertex operators
of the bosonic string. $\Psi[k(s)]$ is a wave functional that describes
a particular state of the string. $k^{\mu}(s)$ is a generalized momentum
and can be expanded as (suppressing the Lorentz index):
\[
k(s) = \ko + {\ki \over s}+ {\kt \over s^2} +...  
\]
\be
=\sum _{n \ge0}k_n s^{-n}
\ee
Similarly one can Taylor expand $\p _z X(z+s)$ as
\[
\p X(z+s) = \sum _{n>0} s^{n-1} {\p _z^nX(z)\over (n-1)!}  
\]   
\be
\equiv \sum _{n>0} s^{n-1} \tY _n(z)
\ee
Thus ($a=1$),
\be
\lp = e^{i\sum _{n\ge 0}\kn \tY _n }
\ee
We use the notation $\tY _0 = X$.
For the bosonic string, one expects $\mu$ to run from 0 to 25. However
we shall let it run from 0 to 26. We will use the 27th coordinate
as the equivalent of the bosonized ghost coordinate, necessary for
representing all the auxiliary fields in the covariant representation
of the string fields \cite{SZ}. We are not going to identify it with the
 ghost coordinate
itself because there is no need to do so and also because we do not wish to be  
forced into any particular representation. The fields are all taken to be massless
in 27 dimensions. Dimensional reduction to 26 dimensions is then required.
$\ko ^{26}$ will be set equal to the mass of the field, in
the free equation. In the interacting equation the prescription will be 
given below. This reduction is quite different from Kaluza-Klein
reduction.
  
The $\kn$ define space time fields:
\[
\lan \kn ^\mu \ran \equiv \int [\prod _{n>0} d\kn ]
 \Psi [\ko ,\ki ,\kt ,...,\km ,... ] \kn ^\mu = S_n^\mu (\ko )
\]
\be   \label{2.15}
\lan \kn ^\mu ~ \km ^\nu \ran = S_{n,m}^{\mu \nu}(\ko )
\ee 
etc.
In order to make the theory gauge invariant we introduce the einbein $\al (s) $
in the loop variable:
\be \label {2.17}
\lpp 
\ee
with the mode expansion:
\[ \al (s) = \sum _{n \ge 0}\aln s^{-n}\equiv e^{\sum _{n\ge 0} \xn s^{-n}}\]

We set $\al _0 =1$.  For reasons explained in \cite{BSLV,P} one has to 
integrate over all $\al (s)$. We assume that 
$[{\cal D}\al (s)] = [ \prod _n d\xn ]$. 

The $\aln $  obey \[{\p \aln \over \p \xm} = \al _{n-m}\]
Defining
 \[Y = X+ 
\sum _{n>0}\aln \frac{\partial^{n}X}{(n-1)!} \equiv
\sum _{n\ge 0} \alpha _{n} \tilde{Y}_{n} \]
and $\yn = {\p Y \over \p \xn}$,
we see that
\be
\lpp = \gvk
\ee

\subsection{Interacting Theory}

The theory is made interacting by the simple modification of
making everything a function of an additional parameter $t$:

\[
k(s) \rightarrow k(s,t) \]
(Thus $\kn \rightarrow \kn (t) $)
\[
X(z) \rightarrow X(z(t))
\]
The parameter `$t$' is only a label for the vertex operator. There is no
functional dependence on $t$. It only enters when we take expectation values
$\lan ... \ran$ (see (\ref{2.214}) below).
We do not want to do this for the $\al (s)$ because the theory
does not possess such a large gauge invariance. In order
to make the choice unambiguous we will translate all $X$'s to
$z=0$ and introduce the einbein there.

Thus we first write
\be
\sum _{n\ge 0}\kn (t) \tY _n (z(t)) = \sum _{n\ge 0}{\bar \kn (t,-z(t))} \tY _n (0)
\ee
This defines $\bar \kn (-z(t))$ to be
\be   \label{2.29}
{\bar k_q (-z)} = \sum _{n=0}^{n=q} k_q  D_{n}^{q}z^{q-n}
\ee
where
\br
D_n^q &  = & ^{q-1}C_{n-1},\; \; n,q\ge 1 \nonumber \\
      &  = & {1\over q}, \; \; n=0 \nonumber \\
      &  = & 1 ,   \;\;           n=q=0
\er

Now we can write the gauge invariant loop variable 
\footnote{These variables $\kn (-z)$ are related to corresponding variables
used in \cite{BSLV2,BSWF}, but the relation  involves $\aln$. Thus
when we treat $\kn (-z)$ as independent variables, this implies a change
of variables. In terms of space-time fields this is a fairly complicated
field transformation.}  
 as
\be
e^{\sum _{n\ge 0}i{\bar \kn (t,-z)} \yn (0)}
\ee

One can also rewrite this as a loop variable analogous to (\ref{2.17}).
Define first, $k(s-z)=\sum _{n\ge 0} \kn (-z) s^{-n}$. Consider
\[
\sum _{n>0}\kn (-z) \tY _n (0)  + \ko X(z) = 
\sum _{n>0}(\kn (-z) + \ko {z^n\over n})\tY _n(0)+ \ko \tY (0)
\]
(The variable in brackets is in fact $\bar \kn (-z)$ defined earlier in (\ref{2.29}.)
\[
=\sum _{n>0}\bar{\kn (-z) } \tY _n(0) + \ko X(0)
\]
\[
=\int ds ~\sum _{n>0}{\bar \kn (-z) }s^{-n} \p X(s) + \ko X(0)
\]
\be
=\int ds {\bar k(s,-z)}\p X(s) + \ko X(0)
\ee
This equation defines $\bar k(s,-z)$.
We can now introduce an einbein $\al (s)$ to get
\be   \label{2.213}
\int ds {\bar k(s,-z)}\al (s) \p X(s) + \ko X(0)
\ee
This should be compare with (\ref{2.17}) of the free theory.

The definition of space-time fields is analogous to (\ref{2.15})
(we will write $z_i$ for $z(t_i)$),
 
\[
\lan \kn ^\mu (t,-z)\ran \equiv \int [\prod _{n>0,t} d\kn (t) ]
 \Psi [\kn (t) ] \kn ^\mu(t) = S_n^\mu (\ko )
\]
\be   \label{2.214}
\lan \kn (t_1,-z_1)^\mu \km (t_2,-z_2)^\nu \ran =
 S_{n,m}^{\mu \nu} (\ko )\delta (t_1-t_2)+ S_n^\mu (\ko (t_1))
S_m^\nu (\ko (t_2))
\ee
Thus when $t_1=t_2$ it describes a higher excitation of one string,
 and when $t_1\ne t_2$ it describes two string modes
interacting.

One can also, if one wants, simplify the notation somewhat by setting $z(t)~=~t$.
In the open string $z$ is real, so this is allowed.  This was done in \cite{BSWF}.

\section{Gauge Transformation}

\subsection{Free Theory}
The gauge transformation in the free case is given by \cite{BSLV},
\be   \label{2.313}
k(s) \rightarrow k(s)\la (s)
\ee
Here $\la (s)$ is a gauge transformation parameter with an expansion
\be
\la (s) = \sum _{n\ge 0} \la _n s^{-n}
\ee
We set $\la _0=1$.

In terms of modes we get:
\be   \label{2.315}
\kn \rightarrow \kn + \sum _{p=1}^n\la _p k_{n-p}
\ee

In order to translate (\ref{2.315}) into space-time fields we will
assume that $\Psi [k(s), \la (s)]$ is also a functional of $\la (s)$.

Thus taking $\lan ...\ran$ on both sides of the equation one gets:
\be
S_n ^\mu (\ko )\rightarrow \ko ^\mu \Lambda _n (\ko ) + \sum _{p=1}^n \Lambda _{p,n-p} ^\mu (\ko )
\ee
where we have set
\[
\lan \la _n \km ^\mu \ran = \Lambda _{n,m}^\mu (\ko )
\]

Note that the photon is $S_1^\mu$ in the above notation and has
the usual Abelian gauge transformation. We will denote it by $A^\mu$ hereafter. 

\subsection{Interacting Theory}

In the interacting case a simple generalization of (\ref{2.313})
gives the following
\be
{\bar k(s,t,-z(t))} \rightarrow \int dt' \la (s,t') {\bar k(s,t, -z(t))}
\ee 
This is very similar to what was suggested in \cite{BSLV0}. However, there
the $k$'s were not $z$-dependent, and consequently only a subset of the interactions
were obtained.
 
In terms of modes:
\be
{\bar \kn (t,-z(t))} \rightarrow {\bar \kn (t,-z(t))} + 
\int dt' \sum _{p=1}^{n} {\bar k _{n-p}(t, -z(t))} \la _p(t')
\ee

To translate this to space-time fields one simply takes expectation
values on both sides. Since the LHS involves, in general, many 
space-time fields, one has to recursively calculate the gauge
transformations of higher level fields after fixing the gauge
transformations of all the lower ones.

The $z_i$'s are variables of integration, and in any
term in the equation of motion they are integrated over a fixed
range. So these integrals are understood on both sides of any
equation.

\section{Equations of Motion}

\subsection{Free Theory}

One first defines the analogue of the Liouville mode. The Polyakov
functional integral  defines the two dimensional conformal field theory.
The  two point function is
\be   \label{1.21}
<X(z) X(w)> \approx {1\over 2} ln ~((z-w)^2 + \eps ^2)
\ee
where $\eps$ is a world sheet cutoff.
On a world sheet where the Liouville mode is $\sigma$ one can let
$\eps \rightarrow \eps e^\sigma$. Thus, when $z=w$,
\be    \label{1.22}
< X(z) X(z)> \approx ln ~\eps + \sigma (z)
\ee

By analogy with (\ref{1.22}) we define
\be  \label{1.23}
\Sigma (z) = <Y(z) Y(z)>
\ee 
neglecting the $ln ~ \eps$ piece. This piece will be retained in the interacting
case. (An alternative way of defining $\Sigma$ is given in \cite{BSLV1}.)
 $\Sigma$ is a function
of $\sigma$ and also $\aln $. When $\al (s) =1$, $\Sigma$ reduces to
$\sigma$.  

In the RG approach, the equations of motion are obtained by requiring the vanishing of
all anomalous $\sigma$ dependences. In the present case we require the
vanishing of $\Sigma$ dependence. Thus consider
\be    \label{1.24}
e^A = :e^{i\sum _{n\ge 0} \kn \yn + \ko .\ko <YY> + 2\sum _{n>0}\ko .\kn <\yn  Y>
+ \sum _{n,m >0}\kn .\km <\yn \ym >}:
\ee
The contractions are the result of normal ordering.
Use 
\[<\yn (z) Y(z)>={1\over2}{\p \Sigma (z)\over \p \xn}\] 
\be  \label{1.25}
<\yn (z) \ym (z) > =
 {1\over2 }({\pp \Sigma (z) \over \p \xn \p \xm}-{\p \Sigma (z) \over \p x_{n+m}})
\ee
To get
\be   \label{1.27}
e^A = e^{i\sum _{n\ge 0} \kn \yn (z) + \ko .\ko \Sigma (z) +
 2\sum _{n>0}\ko .\kn{1\over2}{\p \Sigma (z)\over \p \xn} 
+ \sum _{n,m >0}\kn .\km {1\over2 }
({\pp \Sigma (z) \over \p \xn \p \xm}-{\p \Sigma (z) \over \p x_{n+m}}) }
\ee

The equations of motion are simply obtained by \cite{BSLV}
\be   \label{1.28}
({\delta \over \delta \Sigma }e^A )|_{\Sigma =0}=0
\ee
where integration by parts on all the $\xn$ are allowed.

\subsection{Interacting Theory}

In the interacting case we define 
\be
<Y(0)  Y(0)> = G_\eps (0) + \Sigma (0) 
\ee

Here $G_\eps$ is the coincident two point function of $Y$ on the 
flat world sheet. It is 
a function of $\eps$ and also $\aln$. It reduces to $ln \eps$ when $\al (s) =1$.
\footnote{This is just one convenient choice. Any other ultraviolet cutoff
propagator would do just as well.} 
It is crucial in everything we do, that $\eps$ is {\em finite and non-zero}. 
Off-shell description
of string theory requires this. Otherwise we get singularities. We can only
take $\eps$ to zero in on-shell amplitude calculations.
This equation is a simple generalization
of (\ref{1.22}).  We then replace $\Sigma$ in (\ref{1.27}) by $G+\Sigma$. Also
replace $\kn$ by $\bar \kn (t,-z(t))$. With these replacements eqn (\ref{1.28})
gives the equations of motion. Note that in the free case we did not include $G_\eps (0)$.
This would introduce terms in the equation with different powers of $\eps$. However
in the free theory different powers of $\eps$ are not mixed by gauge transformations,
so they can be safely set to zero. However in the interacting case 
they will be retained.

\section{Gauge Invariance}
\subsection{Free Theory}

A simple way to understand gauge invariance
 is to note that the gauge transformation
(\ref{2.313}) can be compensated by an inverse scaling of $\al (s)$. But
since we are integrating over all $\al (s)$ (equivalently all $\xn$), this
does not affect the functional integral. This assumes that the measure
is invariant. This is true because if we write $\la (s) = e^{\sum _n y_n s^{-n}}$
the gauge transformation simply translates all the $\xn$ by an amount $y_n$
which leaves the measure invariant.
This is equivalent to saying that $A$ changes by a total derivative of the form
${\p \over \p x_p} C$ under
gauge transformation by $\la _p$. 

Thus if $\delta A = \p C =\p (f(\Sigma) B)= (\p f)B + f \p B$, and we vary w.r.t.
$\Sigma$, we get, on integrating by parts, 
$- f'(\Sigma ) \p B + f'(\Sigma ) \p B =0$.

 However there is a subtlety. $\Sigma$ satisfies some
constraints (arising from its definition) and is not a completely unconstrained.
In fact these constraints have to be used to prove that $A$ changes by a total
derivative. 

If one studies the exact expression for $A$ it is easy to see \cite {BSLV3} that if we do
not assume any special properties for $\Sigma$, there are some terms
proportional to $\la _p \kn .\km $ (with both $n,m >0$) that have to be set to zero
if $A$ is to change by a total derivative. Thus we will impose these constraints
on the gauge parameters. These are the familiar ``tracelessness'' constraints
for higher spin gauge fields. Thus we conclude that if we impose these constraints 
the equations are gauge invariant.

\subsection{Interacting Theory}
The structure of $A$ in the interacting case being exactly the same
upto the replacements given above 
(i.e. $\kn \rightarrow \bar \kn (t,-z(t)),
\Sigma (0) \rightarrow G_\eps (0) + \Sigma (0)$) and the form of the gauge
transformations also being the same, the arguments for gauge invariance given above 
for the free case, 
go through here also. The only change is that the constraints have the form
$ \la _p (t) {\bar \kn (t_1 , -z_1)}.
{\bar \km (t_2 ,-z_2)} =0$. Integrations over all variables, $t_i,z_i$, are understood
($t_i$ can be integrated from 0-1, $z_i$ from $0-a$).

\section{Dimensional Reduction}
\subsection{Free Theory}

This was described in \cite{BSLV}. Let us denote the 27th dimension by the
index `$V$'. We simply set $\ko ^V = m$, the mass of the state. 
The kinetic term 
$\ko ^\mu k _{0\mu} \rightarrow \ko ^\mu {\ko}_\mu + \ko ^V k _{0V}$.
Here  $\mu$ runs from $0-26$ on the LHS and $0-25$ on the RHS.

 The
gauge transformation law for $\kn ^V$ under $\la _n$
remains 
\be \kn ^V \rightarrow \kn ^V + \ko ^V \la _n = \kn ^V + \sqrt {n-1} \la _n.
\ee
At the free level all the fields belong to the same level i.e. they have the
same value of $\ko ^V$. There is no inconsistency in setting $\ko ^V$ to
a particular value since gauge
invariance is maintained.

Also it was shown in \cite{SZ} that in order to get the right number
of auxiliary fields the first oscillator of the bosonized ghost has to be
set to zero. This counting was implemented in \cite{BSLV,BSLV0} by imposing constraints
that related terms involving $\ki ^V$ to terms that didn't involve it.
Thus for instance.
\[ \ki ^V \ki ^\mu = \kt ^\mu \ko ^V\]
\be    \label{6.130}
\ki ^V \ki ^V = \kt ^V \ko ^V 
\ee 

The basic idea is to find combinations such that the gauge transformations match.
Of course one also has to find suitable identifications for gauge parameters
of the form $\ki ^V \la _n$ etc. For instance, it is easy to see \cite{BSLV,BSLV0}
that 
\be 
\ki ^V \la _1 = \la _2 \ko ^V 
\ee
is required for the consistency of the identification (\ref{6.130}).
These combinations can then be eliminated from the equations of motion.

\subsection{Interacting Theory}
In order to make contact with string theory it is clear that reproducing 
the Veneziano
amplitude, which involves integrals of the form $(z-w)^{p.q}$, where $p,q$
are 26 dimensional and {\em not} 27 dimensional, requires that
\[
<Y ^V(z) Y^V (w)> =0   ~~~~z\ne w
\]

We also need
\be   \label{6.232}
<Y^V(z) Y^V(z)> = \Sigma (z) 
\ee

In order to implement both of the above we will simply assume
 that $Y^V(z)= Y^V (0)$ and is not a function of $z$.
Thus $\bar \kn ^V(t, -z(t))= \bar \kn ^V (t,0)$.
Since we do not have Lorentz invariance in the 27 th dimension we are free to do
this while retaining the earlier expression for $\mu :0-25$. Note that on-shell
scattering of physical states is not affected by anything we do to $Y^V$.

$\ko ^V $ ($= \sum _{i=1}^N \ko ^V(t_i) $) , the total momentum in the 27th dimension,
in any given term involving $\prod _{i=1}^N k_{n_i}(t_i) $, will be set equal to
$\sqrt {\sum _{i=1}^N n_i -1}$. This counts the number of powers of the cutoff
and is equal to the dimension of the term in the sense of 2-d conformal field theory.
This guarantees that the coefficient
 $\ko .\ko (= \kom . \kom + \ko ^V \ko ^V)$, of $\Sigma$ is nothing but
the RG-scaling operator $\eps {d \over d \eps}$. 
We see this as follows:

In the coefficient of $\ko .\ko \Sigma$, powers of $\eps$ can come from the following
sources:

 1) terms of the form \[e^{p.q G_\eps (z-w)}= e^{p.q ln ~ (\eps ^2 + (z-w)^2)}.\]
$p,q$ being {\em 26-dimensional} momenta.
If we expand in powers of $\eps$ we get $(\eps )^{p.q}$ as well as all powers of
${z\over \eps} , {w\over \eps}$.

 2) terms of the form 
\[\kn ^\mu .\km ^\mu ({\pp \over \p \xn \p \xm} -{\p \over \p x_{n+m}})G_\eps (z-w)\] 
$\mu$ goes from $0-25$.

1) and 2) are responsible for the $z$-dependence of the $\kn ^\mu (-z)$ ($\mu :0-25$).

3) The uncontracted $\yn$ has scaling dimension $n$. This can be made
explicit by measuring all $z$'s in units of $\eps$. Thus we can write
$\yn (z) = \eps ^{-n} \yn ({z\over \eps})$. The overall power of
$\eps$ can then be assigned to $\kn$ that multiplies $\yn$. 
Thus $\kn (-z)$ collects all terms with a given scaling dimension $n$,
in the RG equation.

The expression $\kom .\kom + \ko ^V \ko ^V$ counts all the powers of
$\eps$ described in 1),2) and 3) above.
 
We also need some convention for assigning values to the individual
momenta in the $V$ direction. We will simply assume that every field in an 
interaction term in the equation of motion has equal amounts of it.

The argument that these equations are physically equivalent to those 
obtained from a scattering amplitude calculation can now be made in three
steps. 

First, only the term in the equation of motion coming from $\ko .\ko. \Sigma$
multipliying anything else, is relevant to the scattering of physical states.
All the other terms obtained from derivatives of $\Sigma$
are necessary only for gauge invariance. So we can set them to zero for the
purposes of this argument, and recover them at the end in a unique way
because the formalism is gauge invariant.

Second, set $\al (s)=1$.  Put the $z$'s back into the $\tY$'s (thus
undoing the Taylor expansion) put everything on-shell and take the limit
$\eps \rightarrow 0$. We get
an RG Callan-Symanzik $\beta$-function equation for the coefficient of 
a marginal vertex 
operator. This, by
the usual arguments (see for eg \cite{BSPT,P}), are equivalent to
the scattering amplitudes of string states to all orders. See also
\cite{BSLV2} for some explicit calculations.

Third, now do the Taylor expansion with finite $\eps$ as described in this paper. 
(Removing the $z$-dependence of $\yn$ and putting it into $\kn$ amounts 
to a Taylor expansion.) We have shown that
we get again the RG equations, but now in their Wilsonian form involving
not only marginal but all irrelevant operators. The ``Magic of the Renormalization
Group'' in field theory ensures that when we solve for the irrelevant couplings
and get an equation for the marginal ones, in the limit of going to the infrared
limit, we are guaranteed to get the $\beta$-function calculated directly using
Feynman diagrams. 

Thus the equations obtained here are physically equivalent to those
obtained from string amplitudes.

\section{Tachyon-Photon System}

The tachyon can be included by the simple device of adding to the loop variable
a term $\int dt J(t)$, with the rules
\[
\lan J(t) \ran = \phi (\ko )
\]
\be
\lan J(t_1) J(t_2) \ran = \phi (\ko(t_1)) \phi (\ko (t_2))
\ee
and so on.

The equations are obtained by the following steps:

{\bf Step 1:}

We first write down terms coming from evaluation (\ref{1.28}) that are
proportional to $Y_1^\mu$. We keep terms upto level three, i.e.
involving $\ki \ki \ki , \ki \kt $, and $ k_3$. We find the following terms:
(There is an overall factor of $(\eps ^2 )^{\ko ^2}$ multiplying every term, where
$\ko$ is the total 26-dimensional momentum in any given term):

Level 1:
\[ (\ko (t_1).\ko (t_2)i\bar {\kim} (t_3,-z_3) - \bar \ki (t_1, -z_1).\ko (t_2) i\kom (t_3))(1+J(t_4)) \]

Level 3:
\[\]
a) \[ (-{4\over \eps ^2}) \bar \ki (t_1, -z_1).\ko (t_2) \bar \kt (t_3,-z_3).\ko (t_4)i\kom (t_5) (1+J(t_6)) \]
b) \[ ({2\over \eps ^2})\bar \ki (t_1,-z_1).\ko (t_2) \bar \ki (t_3,-z_3).\bar \ki (t_4,-z_4)i\kom (t_5) (1+J(t_6))\]
c) \be   \label{7.034}
{4\over \eps ^2} \bar \kt (t_1,-z_1).\bar \ki (t_2, -z_2) \ko (t_3).\ko (t_4) i\kom (t_5) (1+J(t_6))
\ee

Gauge invariance is easy to check. The gauge parameters obey the constraint
$\la _1 (t) \bar \ki (t_1,-z_1) .\bar \ki (t_2,-z_2)(1+J(t_3)) =0$ and this
has to be used while checking gauge invariance.

{\bf Step 2:}

We dimensionally reduce by setting (when evaluating $\lan ..\ran$ to convert to space-time fields)
 $\ko ^V= 0$ in the level-1 terms, and also
$\ki ^V =0$. This is consistent since it's gauge transformation involves $\ko ^V$.
In the level-3 terms we set $\ko ^V =\sqrt 2$ for the total momentum. The individual momenta
are equal fractions of this.

Before converting to space-time fields we 
rewrite terms involving $\ki ^V$ in terms of other variables. The following identifications
preserve gauge invariance \cite{BSLV0}:
(The $z$-dependences have been suppressed. Note that because of (\ref{6.232}) $\kn ^V$ has no
$z$-dependence.)

\[2 \bar \ki ^V \bar \kim\bar {\kin} = (\bar {\kim}\bar {\ktn} + \bar {\kin} \bar {\ktm} )\ko ^V\]
\be 2 \la _1 {\ki} ^V \bar {\kim}= (\la _2 \bar {\kim}+ \la _1 \bar {\ktm} ) \ko ^V\ee
 
\[ \bar \ki ^V \bar {\ktm} = 2 \bar k_3 ^\mu \ko ^V - \bar \kt ^V \bar {\kim}\]
\[ \bar {\ki}^V \bar {\ki}^V \bar {\kim}= \bar k_3 ^\mu (\ko^V)^2 \]
\[ \bar {\ki}^V \bar {\ki}^V \bar {\ki}^V = \ko ^V \bar \kt ^V \bar {\ki}^V = k_3 ^V (\ko ^V)^2 \]
\be   \label{subs} 
\la _1 \bar {\ki}^V \bar {\ki}^V = \la _1 \bar \kt ^V \ko ^V = 
\la _2 \bar {\ki}^V \ko ^V = \la _3 (\ko ^V)^2
\ee

{\bf Step 3:}

We convert to space time fields by taking expectaion values $\lan ...\ran$.

{\bf Step 4:}

Gauge transformation of space time fields is determined at each level
recursively, using the transformation of lower levels  as inputs as explained
in Section 3. For instance, the combination 
$\bar \kim \bar \kin \bar \ki ^\rho (1+J(t_4))$ is used
to determine the gauge transformation law of $S_{111}^{\mu \nu \rho}$. As inputs
we use the previously determined laws of $S_{11}^{\mu \nu}$ and
 $S_1 ^\mu (\equiv A^\mu)$ (photon). 
Since it is the same combination that occurs in the equations of motion
(with contracted indices or multiplied by momenta) gauge invariance
of the equations at the loop variable level guarantees invariance at the 
level of space-time fields also. If one looks at the space-time field equations
and their unwieldy gauge transformation laws, this invariance is far from obvious,
though of course it does hold.

As an illustration we write out some of the terms coming from
 Level-1 and Level-3 (\ref{7.034}).
 
{\bf Level-1:} 

\be 
(\eps ^2) ^{\ko ^2}[(k^2iA ^\mu (k) - A(k).k ik^\mu ) + (p+r)^2 iA^\mu (p) \phi (r) - 
A(p).(p+r)i(p+r)^\mu \phi (r)]
 \ee

The gauge transformation of the photon is 
\be
\delta A^\mu(k) = k^\mu \Lambda _1 (k) + r^\mu \Lambda _1 (p) \phi (r)
\ee
Integration over momenta and momentum conserving $\delta$-functions are understood
in all expressions. Thus $p+r=k$ in the above expressions. Also, in the above expression
$\ko ^V =0 = A^V$. Thus the indices run over 26 dimensions only.

By inserting an arbitrary number of tachyons one sees that the gauge transformation 
of the photon can be expressed as follows:
\be
\delta A^\mu (X) e^{\phi (X)} = \p ^\mu (\Lambda (X) e^{\phi (X)})
\ee
Thus the photon field with the canonical transformation law is $Ae^\phi$.
If the normalization of the photon $A$ is fixed due to interactions with the closed string
sector, then at the tachyon minimum, $\phi = -\infty$, the canonical photon field
becomes zero. This supports the arguments for the absence of open string
excitations in the closed string vacuum \cite{GS,AS}.  

{\bf Level-3 :}

On making the substitutions given in (\ref{subs}) we find 

{\bf a)}
\be
(\eps ^2)^{\ko ^2}(-{4\over \eps ^2})[[\bar \ki .\ko \bar \kt .\ko + 2 \bar k_3 .\ko (\ko ^V)^2 +
 k_3^V (\ko ^V)^3] (1+J)]i\kom
\ee

{\bf b)}
\be
(\eps ^2)^{\ko ^2}
({2\over \eps^2})[[ \bar \ki.\ko  + 2 (\ko ^V)^2\bar \ki .\bar \kt + (\ko ^V)^3 k_3^V] (1+J)]i\kom
\ee

{\bf c)}
\be
(\eps ^2)^{\ko ^2}
({4\over \eps ^2})[[\bar \kt .\bar \ki (\ko ^2 + (\ko ^V)^2) + \ko ^V k_3 ^V (\ko ^2 + (\ko ^V)^2)] (1+J)]i\kom
\ee

The next step is to convert to space-time fields by taking expectation values using
(\ref{2.214}).

We will illustrate this on  some of the terms in Level-3 a). The others can be done similarly.
 
The leading $z$-independent term of Level-3 a) gives 
 (converting to position space):
\be
(\eps ^2)^{\ko ^2}
(-{4\over \eps ^2})\{ (-i)\p ^\mu \p ^\rho \p ^\sigma 
[(1+\phi)(S_{2,1}^{\rho \sigma}+ S_2^\rho A^\sigma )] 
+4i\p ^\mu [i\p ^\rho [S_3^\rho (1+\phi)] + 2\sqrt 2 \p ^\mu (S_3 (1+\phi )) 
\ee
We have set $(\ko ^V)^2=2$.

As another example let us look at the $O(z^2)$ piece in $\bar \ki .\ko \bar \ki .\bar \ki$. 
Writing out the $z$-dependence gives:

\be
(\eps ^2)^{\ko ^2}(-{4\over \eps ^2})
(\ki (t_1) +z_1 \ko (t_1) ).\ko (t_2) (\kt (t_3) + z_3 \ki (t_3) +
 {z_3^2\over 2} \ko (t_3) )\ko (t_4) i\kom (t_5) (1+J(t_6))
\ee

Note that $z_i$ are all variables of integration 
and are all being integrated over the same range.
Thus the following are true: $\int dz_i (z_i)^n = \int dz_j (z_j)^n$,
$\int dz_i dz_j (z_i^2 - z_i z_j) = \int dz_i dz_j (-{(z_i-z_j)^2\over 2})$.
Using these and (\ref{2.214}), we find for the $O(z^2)$ piece
\[
 (\eps ^2)^{\ko ^2}\{-6{z^2\over \eps ^2} \mup \pp \p ^\rho [A^\rho (1+\phi )] +2
 {(z -z')^2\over \eps ^2}\mup \p ^\rho \p ^\sigma [A^\rho \p ^\sigma \phi ]\}
\]

The rest of the terms can similarly be evaluated using the same techniques. 
We do not give the  expressions
since they are quite long and not particularly illuminating.

 Note that the $O(z^2)$ term is one of the higher 
derivative terms in the tachyon-photon interaction.
The Koba-Nielsen variables, $z$, are understood to be integrated
over some well defined range, say, $0-a$. Thus the final answers will have in them
a dimensionless number $a\over \eps$ (after a suitable rescaling of the $\kn$).
 This number is a free parameter and is analogous
to the level expansion parameter $4\over 3\sqrt 3$ in BRST string field theory.
It is a measure of how irrelevant an irrelevant operator really is. One can
also set $a=1$ and then effectively $\eps$ becomes that parameter.

What is also noteworthy is that the massive modes
$S_{1,1,1},S_{2,1}$ contribute in a non trivial way to the Abelian gauge invariance 
(whose parameter is $\Lambda _1$). This is clear from the gauge transformation laws
 given below.
As was pointed out in \cite{BSFC} this is due to the finite cutoff on the world sheet.

The gauge transformation law for the fields are given below:

\br
\delta S_{1,1}^{\mu \nu}& = & \Lambda _{1,1}^{(\mu }(k)k^{\nu )} + 
                        \Lambda _1 (p) A^{(\nu }(q) q^{\mu )} +
 r^{(\mu} \Lambda _{1,1}^{\nu )}(p) \phi (r) \nonumber \\
\delta S_2 ^\mu         & = & \Lambda _{1,1}^\mu + 
 \Lambda _1 (p) A^\mu (q) \nonumber \\
\delta S_{2,1}^{\mu \nu}(k) & = & \Lambda _{1,1,1}^{\mu \nu}(k) + \Lambda_{1,2}^\mu (k) k^\nu +
\Lambda _{2,1}^\nu (k) k^\mu +\nonumber \\
&   & \Lambda _1 (p) S_{1,1}^{\mu \nu }(q) +
\Lambda _1 (p) q^\nu S_2 ^\mu (q) + \Lambda _{1,1}^\nu (p)A^\mu (q) + 
q^\mu \Lambda _2 (p) A^\nu (q) \nonumber \\
&   & +\Lambda _{1,2}^\mu (p) r^\nu \phi (r) - 
{(z_1-z_3)^2\over 2}p^\nu r^\mu \Lambda _1 (p) \phi (r) \nonumber \\
\delta S_{1,1,1}^{\mu \nu \rho} & = & k^{(\mu }\Lambda _{1,1,1}^{\nu \rho )} +
 \Lambda _1(p) q^{( \mu } S_{1,1}^{\nu \rho )}(q)
\nonumber \\
&   &+ [[\Lambda _{1,1}^\nu (p) A ^\rho (q) + (\rho \leftrightarrow \nu )]q^\mu 
+~ two ~permutations~] + 
\nonumber \\
&   & z [\Lambda _1(p) q^\mu (p+q)^\nu A^\rho + (\rho \leftrightarrow \nu ) +
 two ~ permutations~]+ r^{(\mu}\Lambda _{1,1}^{\nu \rho )} \phi (r)
\nonumber \\ 
&   & 
- z \Lambda _1 (p) \phi (r) A^\nu (q) [(r^\mu (p+q)^\rho + r^\rho (p+q)^\mu + r^\mu r^\rho )
~+~two~ permutations~]\nonumber \\
&   & - (z_4-z)^2 \Lambda _1(p) \phi (r) r^{( \mu} r^\nu p^{\rho )} \nonumber \\
\delta S_3^\mu & = & \Lambda_{1,2}^\mu (k) + \Lambda _{2,1}^\mu (k) + 
\Lambda _3 (k) k^\mu + \Lambda _1(p) S_2^\mu (q) + \Lambda _2 (p) A^\mu (q) \nonumber \\
&   &+
z \Lambda _{1,1}^\mu -z \Lambda _1 (p) A^\mu (q) - z k^\mu \Lambda _2(k)  -
{z^2 \over 2} k^\mu \Lambda _1 (k) - {z^2\over 2}\Lambda_1(p)r^\mu \phi (r) .
\nonumber \\
\delta S_2 & = & 2\Lambda _2 \nonumber \\
\delta S_{2,1}^\mu & = & {3\over \sqrt 2} \Lambda _{2,1}^\mu (k) +
({3\over \sqrt 2}-2)\Lambda _2 (p) A^\mu (q) - S_2(p)q^\mu \Lambda _1 (q) +
\nonumber \\
&   & {1\over \sqrt 2} \Lambda _{1,2}^\mu + {1\over \sqrt 2} \Lambda _1 (p) S_2 ^\mu (q)  + 
\sqrt 2 \Lambda _3 (k) k^\mu +\nonumber \\
&   & ({3\over \sqrt 2}-2) z\Lambda _2 (k) k^\mu + {1\over \sqrt 2} z \Lambda _{1,1}^\mu +
 {1\over \sqrt 2} z \Lambda _1 (p) A^\mu (q) \nonumber \\
&   &+ {z^2\over 2\sqrt 2} k^\mu \Lambda _1 (k) + \sqrt 2 \Lambda _3 (p) r^\mu \phi (r) +
({3\over \sqrt 2}-2)z \Lambda _2 (p) r^\mu \phi (r)
 +{z^2\over 2\sqrt 2}\Lambda _1 (p) r^\mu \phi (r). \nonumber \\
\delta S_3 & = & 3\sqrt 2 \Lambda _3 .
\er

The gauge invariance of the equations of motion are much easier to verify
before dimensional reduction. We have done this explicitly for some of the
gauge invariances: $\Lambda _{1,1,1}$ at $O(z^0)$, $\Lambda_{1,1}$ at $O(z)$ 
 and $ \Lambda _1 \phi$ at $O(z^2)$.  As explained earlier this
follows necessarily from the gauge invariance at the loop variable
level.

\section{Summary and Conclusions}

In this paper we have described a solution to the problem of obtaining
gauge invariant equations of motion for the modes of the open bosonic
string using the RG equations of the world sheet conformal field theory.
This approach involves defining variables on a curve (``loop variable'').
In this approach, there are several intriguing features. First, the theory is formally
written as a massless theory in 27 dimensions and masses are obtained by
a dimensional reduction prescription (that is quite a different one from
the usual Kaluza-Klein reduction). Second, the structure
of the interacting theory, both the form of the equations and the gauge 
transformation law, is similar to that of the free theory. The loop is just 
thickened to a band and the loop variables acquire a dependence on the 
positions of the vertex operators. Third, the gauge transformation law,
in terms of loop variables has a simple interpretaion of space-time scale
transformations. This supports the speculation \cite{BSLV} that the  space-time
Renormalization Group on a lattice with finite spacing, is part of the invariance
group of string theory.
Finally,  space-time gauge invariance of the equations obtained this way
does not seem tied down to any special world sheet properties, unlike in BRST
string field theory where it follows from BRST invariance. To that extant
it need not describe a string theory. Only the special choice of Green's function
enforces the string theory connection.

There are many open questions.
We have not investigated the issue of whether there is a simple generalization
that works for loops. The precise relation to BRST string field theory 
is not clear. The theory is so much simpler in terms of loop variables that
it would be interesting to work out solution generating techniques in terms
of these variables rather than in terms of space-time fields.
Finally, it would be interesting to find a physical explanation of the
``intriguing'' features mentioned above.

\end{document}